# Improvement of Key Problems of Software Testing in Quality Assurance

Nayyar Iqbal, M. Rizwan Jameel Qureshi
Department of Computer Science
COMSATS Institute of Information Technology
Lahore Pakistan
nayyar_iqbal@hotmail.com, rjamil@ciitlahore.edu.pk

**Abstract**
Quality assurance makes sure the project will be completed based on the previously approved specifications, standards and functionality. It is required without defects and possible problems. It monitors and tries to progress the development process from the start of the project. Software Quality Assurance (SQA) is the combination of the entire software development process, which includes software design, coding, source code control, code review, change management, configuration management and release management. In this paper we describe the solution for the key problems of software testing in quality assurance. The existing software practices have some problems such as testing practices, attitude of users and culture of organizations. All these tree problems have some combined problems such as shortcuts in testing, reduction in testing time, poor documentation etc. In this paper we are recommending strategies to provide solution of the said problems mentioned above.

## 1. Introduction
Developing a [1] good software system is a very difficult task. To make a good software product, numerous measures for software quality attributes need to be taken into explanation. System complication dimension plays a vital role in controlling and supervision of software quality because it normally affects the software quality attributes like software reliability, software testability and software maintainability. Thus, software quality assurance (SQA) [1] needs to be addressed keeping in view the new strategies, tool, methodologies and techniques applicable to software development life cycle.

Software quality [2] is gaining much more interest these days as well as much more importance is being given to the production of high quality software products. Software development is a intricate process requiring careful integration of various disciplines, technical activities, project management etc. Most software are produced by the joint effort of many designers and programmers working over a period of man years. The resulting product can't be totally understood by any person. No matter how [2] well-designed methods used to test the final product, how complete the documentation, how structured the methodology, the development plans, the project reviews, the walkthroughs, the database management, the configuration control, no matter how advanced the tools and techniques - all will come to nothing and the project will fail if the quality management system is not effective.

## 2. Related Work
In this paper the author [1] describe that insecurely tested software system lowers down the system reliability that afterward negatively affects 'Software Quality'. In this paper 'Software Reliability Measurement' has been discussed and also ISO approach applicable to software quality assurance (SQA). In order to increase the efficiency of testing and to improve software quality, software houses must make transitions to higher software culture. Testing need to concentrate on maximizing customer satisfaction rather than just detecting and correcting errors involved in delivered software. In this paper, the factors affecting software quality management have been discussed and the author suggested possible improvements. The results of this paper may be quite supportive to the researchers in quantifying the specific measuring tools for these software qualities attributes.

In their paper [2] J Barrie Thompson and Helem M Edwards says that given the fact that time is very limited in the course the authors believe that in the Systems Engineering module they have been able to provide an appropriate balance between master's level research activities and those of a more practical nature. However, the authors believe that they have been able to present the students with some really helpful practical experiences. In particular the positive features of the approach the authors have adopted are: The prepared arrangements in particular with consider to the front-loading of formal lectures means that students receive an early overview of the subject area. They then have sufficient time to explore their chosen research area(s) in depth. The practical side of the module provided the students with the chance to put into practice some of the aspects of software engineering that they had encountered and begun to understand from their research activities.

In the [3] modern years an increasing number of software organizations have launched initiatives to improve their software process. The majority of them have been not capable to move beyond diagnosis and action planning, turning those plans into real and practical actions. This paper focuses on two software process areas, Software Quality Assurance (SQA) and Software Configuration Management (SCM), and proposes a set of basic tools to assist in the implantation of specific practices for them. SQUID (software Quality In the Development Process) - for specifying, monitoring and evaluating the software product quality during development is adapted. The authors describe the results of a application conducted, showing how the proposed adaptation helps in formalizing and normalizing the implantation process, setting tangible goals and evaluating the results more accurately.

The authors [4] illustrate that software quality assurance is faced with many challenges starting with the method of defining quality for software. There must be complete understanding what high quality software is, but the final description is generally influenced by the environment of the software usage. There are many aspects of SQA from those within the phases of the software development life cycle to those that span several phases. SQA is a very difficult area that is serious to the final success of a project; it is also one that requires a rather diverse set of skills. New information areas such as software safety and reliability are now being added to the core set of required skills. SQA must be independent from development organizations to be successful.

Massood Towhiddnejad [5] describes an experiment which involved students in the undergraduate computer science senior project capstone design course, and students in the graduate software testing course. Students entering the senior project class are graduating seniors who have completed all but maximum of two required CS classes. They have already completed a one semester software engineering class with major concentration on software development life cycle and software processes. Students entering the software testing course have already completed graduate course in software engineering, project management, requirement engineering, and they may have other courses in software design and architecture. Students in the undergraduates' classes worked as the development team while students in the graduate classes worked as the software quality assurance team, both working on a single product.

The authors [6] have addressed a practical drawback of software metrics based quality classification models based on Boolean Discriminant Functions. More specifically though BDFs have confirmed excellent ability to predict fault prone modules they do so at a very high inspection cost. However it should be noted that there may be situations in which software development organizations are agreeable to deal with relatively high inspection costs, provided all low quality modules are reviewed and enhanced.

This paper applies [7] Lehman's theory of software evolution to analyze the characteristics of web-based applications and identifies the essences and incidents that cause difficulties in developing high quality web-based applications. It is argued that they belong to Lehman's E-type systems, hence satisfy Lehman's eight laws of software evolution. The doubts underlying the development of web applications are analyzed and their implications are discussed. In order to support sustainable long term evolution of such systems, authors proposed a cooperative multi-agent system approach to support both development and maintenance activities. A prototype system with emphasis on testing and quality assurance is reported.

This paper describes [8] the one of the most important things that students can learn in a course in software engineering is how to effectively work in a team to develop software that is too large for a single individual to produce. It is also essential that students learn the value of assuring software quality at each step of the development process. This paper also illustrates how to include a UML-based team project into an object oriented software engineering course. The project gives students practical experience in software development and quality assurance at each stage of the software lifecycle, including analysis, design, implementation, and integration. In this paper author describe approach and example project that includes the problem requirements, timetable of deliverables and sample deliverables.

This paper has [9] shown that the software quality has been advanced, and Software Process Improvement (SPI) has been also improved since the execution of system based SQA. Samsung Electronics Semiconductor Business (SESB) is trying to improve the better software quality and gain optimum process as in many companies. Therefore, SESB constructed IT-Workplace and ITPM to elevate productivity of software development and optimize Software development process. Based on IT-Workplace and IT Product Management, SQA department has conducted the activities of increasing software quality including audit and Software Process Improvement.

The Authors observed [9] that software quality has been increased and stabilized by appraising audit score and total time of software development process in the software project has taken less time than before. In order to maintain good software quality, SQA department constantly has to search for the improvement point, get the opinion from project team, and reflect on that view and improve the process. In summary, authors conclude that the introduction of the system based SQA has a strong effect on the elevation of SQA and SPI.

This paper describes [10] quality assurance (QA) methods such as software testing and peer review are very important to reduce the adverse effects of defects in software engineering. In this paper the authors explore current practices of QA and possibilities for their extension in open source software (OSS) projects. This paper presented a framework for QA aspects in OSS project based on our observation from typical OSS projects. The authors identified major challenges for future work [10]: a) how to better formulate such indicators as the basis of meaningful notifications about the status of OSS product quality for different stakeholders, b) how much effort seems reasonable to spend on creating, maintaining and monitoring the indicators in an OSS context; and c) the need for empirical evaluation of the concept using larger set of OSS projects.

## 3. Hypothesis

Hypothesis of our paper is that we provide strategy for improvement of key problems such as shortcuts in testing, reduction in testing time, let go deliver now correct errors later attitude, poor planning and co-ordination, lack of users involvement, poor documentation, lack of management support, inadequate knowledge of application environment, improper staffing and poor testability. In this paper we are focusing on the above factors to bring improvement in them

## 4. Strategy for improvement of key problems

In this paper we give strategy for the improvement of key problems [1] which are being faced in the software quality assurance during testing.

### 4.1 Shortcuts in testing

Testing is considered tough task by many software project managers and software houses. Software testing is a creative and complicated task required expert, active and energetic employees in software development. Following steps are necessary to avoid from shortcuts in testing**.** Obtain requirements, functional design, and internal design specifications and other necessary documents. We must also obtain schedule requirements that determine project-related personnel and their responsibilities, reporting requirements, required standards and processes such as release processes, change processes, etc. Testing team must identify application's higher-risk aspects, set priorities, and determine scope and limitations of tests. Determine test approaches and methods - unit, integration, functional, system, load, usability tests, etc. Test environment requirements must also be determined such as hardware, software; communications, etc. Determine testware requirements such as record /playback tools, coverage analyzers, test tracking, problem/bug tracking, etc. Test input data requirements that are to be used during testing must be determined. Identify tasks, those responsible for tasks, and labor requirements. In testing process we must set schedule estimates, timelines, and milestones.

Preparation of test plan document is very necessary during testing. Test cases are to be written before starting testing. Prepare test environment and testware, obtain needed user manuals/reference documents/configuration guide/installation guides, set up test tracking processes, set up logging and archiving processes, set up or obtain test input data. Testers must obtain the software developed by the developers and install software to check is there any error. After installing the software performs tests evaluate it and report results. Track problems/bugs and fixes and retest as needed. Maintain and update test plans, test cases, test environment, and testware through life cycle.

### 4.2 Reduction in testing time
In this we must follow the following steps. In order to give appropriate time to testing, software engineers must follow the schedule. Time required for each phase of development must be followed; in reality testing is often estimated inadequately. Design and coding generally take more time than estimated or planned therefore proper management must be done in order to avoid from reduction in testing time.

### 4.3 Let go-deliver now, correct errors later-attitude
Possible areas of improvement include Testing team must completely participative in testing. Each member of testing team must focus on rules of testing as defined by the software house. There must be better planning and effective co-ordination among the testing team and development team. Feedback and search for continuous improvement must be considered among testing team.

### 4.4 Poor planning and co-ordination
**Planning**
Planning for testing must be considered in prior phases of software development, testing is not given appropriate time till the last stages of the project. This checklist should be used at each planning stage. Collect important documents, such as: previous version of the documentation plan, specifications requirements documents, documentation proposal quality plan. Planning is done to the following factors provision of personnel and equipment to be used during the software development must be properly planned. Assign responsibilities for aspects of the documentation. Team leader is required to estimate the financial costs during software development. Preparation of schedules is very necessary at planning phase. Complete planning is required to know which prototypes are to be used when. Documentation reviews are also needed to check the weakness in the previous projects. There must be complete coordination between developers and customer to make project successful and approval mechanism for the documentation. Decide how to handle updates and future developments. Write the documentation plan again if necessary.

**Coordination**
Test team and design team must have complete coordination among them to save the project from any damage. Customer coordination among design and test team is required to be established to give complete satisfaction to the customer.

### 4.5 Lack of user involvement
User plays very important role in testing process. The concepts of joint application design (JAD) and the group support system (GSS) can be used for user involvement and are getting better acceptance in software development. They make possible active, vigorous, interaction between users and developers. The developers need to draw the concentration of users in testing and support their involvement in test planning, system testing and acceptance testing.

### 4.6 Poor documentation
These two types of documentation user documentation and system documentation are very necessary during the development of the software. Improvement must be done to the following factors in order to avoid from poor documentation. Check the missing information throughout. Poor writing and ambiguity always create big problems improvement is required to this factor. A failure to anticipate the reader's problem, questions, and environment. Documents are written for the writers and their environment not for the readers and their environment documents must be suitable for the readers. Focus on the improvement of wrong technical level. Formatting and design of structure plays very important role in documentation. Proper indexing of documentation is also very necessary (documentation contain good information but it is hard to find). A professional appearance which belies the poor content and from failure to match changes to a product. Lack of planning for documentation is also reason for poor documentation.

### 4.7 Lack of management support
Excellence principles can only be achieved by employing effective quality management structure. The management and technical procedures built quality into software product that are defined and implemented to guarantee: quality, schedule and budget compliance. There are various technologies for software upgrading, which includes the most important objective of software engineering. Few examples of the important technologies are [1] requirement definitions, defect prevention, defect detection and defect removal.

### 4.8 Inadequate knowledge of application environment
Testing team must have a complete information of the functionalities of the software being tested, its users and the plateform in which it is going to work without these it causes incorrect focus on testing and missing those area that more significant to user. Important user necessity may be missed without information of the environment.

### 4.9 Improper staffing
Testing is a team work and everyone in the team must work for the achievement of testing. Appointment of correct team member for testing has great control on the achievement of testing. In testing we need team members that have experience of development and in testing. The team leader should have the qualities of problem solving and management skills and capacity to supervise a team and synchronize with clients.

### 4.10 Poor testability
Software testing consists of scheduling, effort and time. Software validation and verification techniques must be used to test software in order to avoid from testability. After development we recommend following validation testing black box testing, white box testing, unit testing, integration testing, system testing and acceptance testing. For verification testing there must be peer and group reviews of software between the customer and developers at various development phase. Software quality assurance activity formal technical review must be made during verification testing. Developers are required to develop the software which has the following characteristics operability, observability, controllability, decomposability, simplicity, stability, and understandability.

## 5. Conclusions
Software testing is the process of executing an implementation of the software with test data and examining the output of the software. In testing software testing techniques, methodologies, tools and principles can be supported. It is the management responsibility for effective testing. Testing team members need to focus on agreement made with customer. In this paper we recommend strategy for improvement of key problems of software testing in quality assurance. Following problems are considered such as shortcuts in testing, reduction in testing time, let go deliver now correct errors later attitude, poor planning and co-ordination, lack of user involvement, poor documentation, lack of management support, inadequate knowledge of application environment, improper staffing, poor testability.